\documentclass[prb,floatfix,twocolumn,amsmath,amssymb,showpacs]{revtex4}
\usepackage{graphicx}

\begin{document}

\title{Numerical study of finite size effects in the one-dimensional
two-impurity Anderson model}

\author{S. Costamagna and J. A. Riera}
\affiliation{Instituto de F\'{\i}sica Rosario, Consejo Nacional de
Investigaciones Cient\'{\i}ficas y T\'ecnicas,\\
Universidad Nacional de Rosario, Rosario, Argentina
}

\date{\today}

\begin{abstract}
We study the two-impurity Anderson model on finite chains using
numerical techniques. We discuss the departure of magnetic 
correlations as a function of the interimpurity distance from a pure
$2k_F$ oscillation due to open boundary conditions. We observe
qualitatively different behaviors in the interimpurity spin correlations
and in transport properties at different values of the impurity
couplings. We relate these different behaviors to a change in the
relative dominance between the Kondo effect and
the Ruderman-Kittel-Kasuya-Yoshida (RKKY) interaction. 
We also observe that when
RKKY dominates there is a definite relation between interimpurity
magnetic correlations and transport properties. In this case,
there is a recovery of $2k_F$ periodicity when the on-site
Coulomb repulsion on the chain is increased at quarter-filling.
The present results could be relevant for electronic
nanodevices implementing a non-local control between two
quantum dots that could be located at variable distance
along a wire.
\end{abstract}


\maketitle

\section{Introduction}
\label{intro}

It is well-known that the main properties of the two-impurity Anderson
model (TIAM)\cite{tiam_history} are determined by the competition 
between the Kondo effect\cite{hewson} generated by each impurity and
the Ruderman-Kittel-Kasuya-Yoshida (RKKY) interaction between the 
impurities.\cite{tiam_main_res,hirsch}
Renewed interest in this model comes from the observation of the 
Kondo effect in electronic devices where a nanoscopic
relatively isolated region, a quantum dot (QD), can act as a single
spin-1/2.\cite{goldhaber} In particular, 
it has been observed in a nanoelectronic device formed by
two QDs coupled through an open conducting region\cite{craig} that 
transport through one of the QDs depends on the occupancy and 
coupling of the other QD via the RKKY interaction. The main issues
which are involved in this and related experiments\cite{heersche}
are the possibility of tuning different regimes dominated by the
Kondo effect or by the RKKY interaction, and the dependence
of transport properties with the sign of the magnetic 
correlation between the impurities.

\begin{figure}
\includegraphics[width=0.43\textwidth]{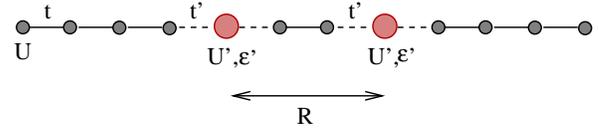}
\caption{(Color online) Picture of model (\ref{ham-TIAM}),
$L=12$, $R=3$.
}
\label{fig1}
\end{figure}

Analytical studies of the
TIAM are usually concerned with long-distance 
behaviors.\cite{tiam_history,tiam_main_res} However, it should
be emphasized that the properties of mesoscopic devices
are determined by finite size effects rather than by bulk physics.
Finite size effects have been addressed in a
study of a double QD in an Aharonov-Bohm ring\cite{utsumi}
and it was found that RKKY dominates the transport properties.
More closely related to the problems we consider in this work
is a study of two impurities coupled by a finite one-dimensional
(1D) wire\cite{simoncloud} where it was shown that the RKKY
interaction is always dominant due to the strong reduction of
the Kondo temperature by finite size effects. In another
slave-boson study\cite{simon} it was concluded that the
presence of different transport regimes depends on the sign 
of the RKKY interaction.
In these two studies however, the impurities are located at the
edges of the chain and hence the effect of the outer sites in
Fig.~\ref{fig1}, or ``leads", were not considered.

The main motivation for the present study is then to 
understand finite size effects in the 1D TIAM which can be
considered as a model of a device formed by two QDs coupled
through a conducting chain and connected to 1D leads 
(Fig.~\ref{fig1}).
Another motivation for this work is to understand the effects of
electronic correlations which are present in materials such
as carbon nanotubes\cite{bockrath} used in experimental
realizations of quasi-1D devices.\cite{biercuk} Coulomb correlations
in the chain affect both the Kondo temperature and RKKY 
interaction.\cite{simoncloud,egger-schoeller}

In this work we show numerically that the spin-spin correlation
between the two impurities as a function of the interimpurity
distance $R$, $S(R)$, for a fixed length $L$ of the whole chain,
departs from the predicted $2k_F$ long-distance behavior for the
TIAM.\cite{tiam_main_res,hirsch} We show how a regime where
the RKKY interaction dominates over the Kondo effect appears as
the impurity couplings are decreased. We also provide a
measure of the conductance $G$ which is more
relevant for applications to nanodevices. We show that this
quantity, particularly in the regimes where RKKY dominates, 
follows an oscillatory behavior with $R$ at a fixed $L$ related
to that of $S(R)$. We also show that an on-site Coulomb 
repulsion on the chain, at quarter-filling, restores the 
$2k_F$ periodicity of $S(R)$ and $G(R)$. Finally, we provide 
indications that most of these finite size effects found are 
due to the open boundary conditions adopted for our system.

The paper is organized as follows. In Sec.~\ref{model} we 
define the Hamiltonian model and we describe the method
employed in the calculations. In Sec.~\ref{results} we 
present results for $S(R)$ and $G(R)$, first for 
noninteracting leads and then in the presence of correlations
on the chain. We conclude in Section~\ref{conclusion} by
providing a summary of the results and by suggesting possible 
connections with other
theoretical approaches as well as with experiments.

\section{Model and methods}
\label{model}

The 1D two-impurity Anderson model with Hubbard repulsion 
on the chain is defined by the Hamiltonian:
\begin{eqnarray}
{\cal H}_0 = &-& t_0 \sum_{i,i+1\in \Lambda,\sigma}
(c^{\dagger}_{i+1 \sigma} c_{i \sigma} + H.c. )
+ U \sum_{i\in \Lambda} n_{i\uparrow} n_{i\downarrow} \nonumber\\
&-& t' \sum_{\mu =\pm1, \sigma} (c^{\dagger}_{r_1+\mu \sigma}
c_{r_1 \sigma} +
c^{\dagger}_{r_2+\mu \sigma} c_{r_2 \sigma} + H.c. )   \nonumber\\
&+& U^\prime \sum_{i=r_1, r_2} n_{i\uparrow} n_{i\downarrow} +
\epsilon^\prime \sum_{i=r_1, r_2} (n_{i\uparrow}+n_{i\downarrow})
\label{ham-TIAM}
\end{eqnarray}
\noindent
where the notation is standard. The Anderson impurities or QDs, with
parameters $U^\prime, \epsilon^\prime$, are symmetrically located
with respect to the center of the chain, i.e. at sites
$r_{1,2}=(L\pm R+1)/2$, with $R$ odd (see Fig.~\ref{fig1}). The
QDs are connected to the rest of the system $\Lambda$
with a hopping $t^\prime$. The subsystem
$\Lambda$ comprises the leads ($i<r_1$, $i>r_2$) and the region
between both impurities ($r_1<i<r_2$), and it is described by a
Hubbard Hamiltonian with couplings $t_0$ and $U$.
$t_0=1$ is adopted as unit of energy.

Model (\ref{ham-TIAM}) will be studied on finite clusters using
density matrix-renormalization group (DMRG).\cite{dmrgrev}
Open boundary conditions (OBC) were adopted throughout. We
would like to emphasize that OBC are the realistic boundary 
conditions for the devices that motivate the present work. There
has been a previous study of a related model using DMRG but the two
impurities were fixed at the chain edges,\cite{hallberg-egger}
that is without leads.

We provide a measure of the response of the system to the
application of a small bias voltage $\Delta V=V_R-V_L$
($V_R=-V_L$), where $V_L$ ($V_R$) are on-site potentials applied
to the ten sites at the edge of the left (right) lead.
Although these 1D leads have to be connected to massive two- or
three-dimensional metallic or semiconducting contacts, we believe
that as a first step it is necessary to consider the system formed
only by the two QDs connected to the 1D leads.
The current $J_L(t)$ ($J_R(t)$) on the bond connecting the left
(right) QD to the left (right) lead is computed with the
time-dependent DMRG.\cite{schollwock} In the case of a single
QD and non-interacting leads, this numerical setup has
been shown\cite{cazalilla,whitefeiguin,alhassanieh,schmitteckert}
to reproduce the results for equivalent systems which were
treated analytically.\cite{wingreen} In the following, we employed
the ``static" algorithm\cite{cazalilla} which although less precise
than the ``adaptive" algorithm\cite{whitefeiguin} still gives
qualitatively correct results, as we have explicitly checked in
some cases,\cite{qdhub,lamnoimp}
but is much faster than the adaptive scheme thus allowing to explore
different couplings and densities and a wide range of $R$ and $L$.
All the results reported below correspond to $\Delta V=0.01$.
We compute $J(t)=(J_L(t)+J_R(t))/2$, which follows with time very
approximately a sinusoidal function, and we adopt as a measure of
the response or ``conductance", $G$, the average of the sinusoidal
fitting to $J(t)/\Delta V$ over a half-period. This qualitative
behavior is mostly independent of various criterions that have
been proposed to quantify $G$.\cite{alhassanieh}
We want to emphasize that our main purpose in this study is to
obtain {\em relative} values of $G$ as a function of the
interimpurity distance $R$. In this sense, the application of
the bias potential on only ten sites at the edges of the
system instead of applying to the whole leads is then a necessary
condition to treat all distances $R$ on an equal foot.

We adopted $U^\prime =8$ and
$\epsilon^\prime =-4$ (symmetric point) in order to ensure a strong 
magnetic character of the impurities which we checked in all cases.
In the following we will consider two values of $t^\prime$, 
$\sqrt{2}/2$ and 1. The Kondo effect for the single impurity Anderson
model for $t^\prime=1$ was studied using the same method as in the
present work in Ref.~\onlinecite{qdhub}.
Most of our calculations reported below were obtained 
for $L=120$. The truncation error in DMRG calculations for this
cluster was kept below $10^{-6}$.
In all cases we worked in the subspace of total $S^z=0$. Thus,
taking into account the isotropy of the Hamiltonian (\ref{ham-TIAM})
we computed the spin-spin correlations between sites $i\neq j$ as
$S_{ij}=3\langle S^z_i S^z_j \rangle$. In particular, the
spin-spin correlation between impurities is
$S(R)=3\langle S^z_{r_1+R} S^z_{r_1} \rangle$.

\begin{figure}
\includegraphics[width=0.43\textwidth]{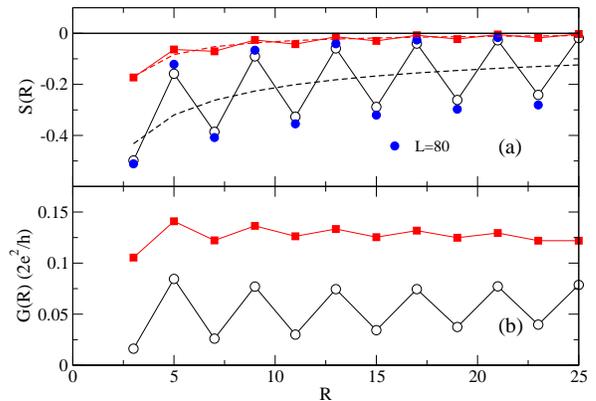}
\caption{(Color online) (a) Spin-spin correlations between the
impurities, and (b) conductance,  as a function of the 
interimpurity distance $R$, at $n=1$, for $t^\prime=\sqrt{2}/2$
(circles) and $t^\prime=1$ (squares), on the $L=120$ chain except
otherwise stated. In (a), dashed lines show fits to $2k_F$
oscillations.}
\label{fig2}
\vspace{-0.3cm}
\end{figure}

\section{Results and discussion}
\label{results}

Let us start by discussing results at half-filling, $n=1$. In
Fig.~\ref{fig2}(a), it can be seen that for $t^\prime=\sqrt{2}/2$
and $t^\prime=1$, $S(R)$ presents a 4-site period which corresponds
to a $k_F$ oscillation. To be more precise there is a $k_F$
modulation on top of the antiferromagnetic (AF) $2k_F$ oscillation.
This $2k_F$ component can be fitted for $R$ odd by the law 
$-1.102/x^{0.59}$ ($t^\prime=\sqrt{2}/2$) and
$-1.021/x^{1.38}$ ($t^\prime=1$). The $k_F$ component decays much
slower, as $x^{-0.22}$ ($t^\prime=\sqrt{2}/2$) and $x^{-0.68}$ 
($t^\prime=1$).  At this point we could
advance the hypothesis that the departure of the present results
from the pure $2k_F$ oscillation is due
to the OBC used in DMRG calculations. That is, $2k_F$ oscillations
starting from the chain edges would modulate the magnetic
correlations from each of the impurities and hence, as a function
of $R$, this would enhance the $k_F$ component of $S(R)$.
The large intensity of $S(R)$ and its survival at long
distances for $t^\prime=\sqrt{2}/2$ suggest that we are in the
regime where the RKKY interaction dominates over the Kondo effect.
On the other hand, the smaller amplitude and rapid suppression of
$S(R)$ with $R$ for $t^\prime=1$ indicate the presence of
a strong screening of the magnetic moment of the impurities
implying a dominance of the Kondo effect. 

The conductance (Fig.~\ref{fig2}(b)) presents an oscillation with $R$
that follows that of $S(R)$. There is a set of minimum values of
$G(R)$ for $R=4 m+3$ ($m$ integer) that correspond to the stronger
AF spin-spin correlations between impurities,
and a set of maximum values of $G(R)$ which occur at $R=4 m+1$
corresponding to the weaker AF $S(R)$. The expected difference
in amplitude of $G(R)$ between both values of $t^\prime$
considered, since the current is proportional to $t^\prime$, can
also be observed. The most important feature in these results is 
that, for the case of $t^\prime=\sqrt{2}/2$, there is a factor
of 2 or larger between the conductance for $R=4 m+1$ with respect
to the one for $R=4 m+3$. We believe that this large difference
can be detected experimentally in appropriate devices. In our 
calculation, the maximum of $G(R)$ could not possible reach the
expected unitary limit as a consequence of applying the bias
potential to only few sites on the leads as discussed in 
Section \ref{model}.

\begin{figure}
\includegraphics[width=0.43\textwidth]{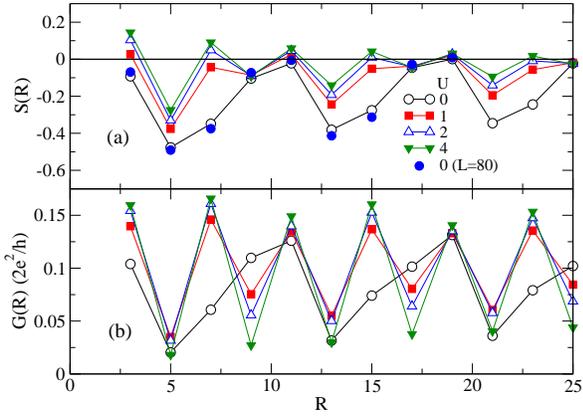}
\caption{(Color online) (a) Spin-spin correlations between the
impurities, and (b) conductance, as a function of the interimpurity
distance $R$, for $t^\prime=\sqrt{2}/2$, at $n=0.5$, parameterized
with $U$, on the $L=120$ chain except otherwise stated.}
\label{fig3}
\vspace{-0.3cm}
\end{figure}

Although half-filling is perhaps more realistic for actual devices,
it is interesting to study also the case of quarter-filling, $n=0.5$.
In this case it is possible to study transport properties in the
presence of electron correlations on $\Lambda$. In
Fig.~\ref{fig3}(a), for $t^\prime=\sqrt{2}/2$, it can be seen that
for the noninteracting chain ($U=0$) $S(R)$ presents an oscillation
with period 8 which again corresponds to an overall $k_F$ oscillation.
As in Fig.~\ref{fig2}(a), results for $L=80$ show
an oscillation in $S(R)$ slightly larger in amplitude to
that for $L=120$. A systematic study of the effect of $L$
is presented at the end of this work.
Similarly to what happened at $n=1$, the
conductance, shown in Fig.~\ref{fig3}(b), has the same dependence
with $R$ as the spin-spin correlations. At $n=0.5$, the
minimum (maximum) of $G$ occurs at $R=8 m+5$ ($R=8 m+3$) which 
coincides with the values of $R$ at which $S(R)$ has the strongest
(weakest) AF values.

Let us turn on the interaction on $\Lambda$. It can be seen in
Fig.~\ref{fig3}(a) that there is a rapid suppression of the 
$k_F$ modulation in $S(R)$ with increasing $U$ leaving behind a 
well-defined $2k_F$ oscillation which is already apparent at
$U=2$. For $U \geq 2$ the spin-spin correlations at $R=4 m+3$
become slightly positive, that is ferromagnetic (FM). Consequently,
as it can be observed in Fig.~\ref{fig3}(b), there is a similar change 
in the dependence of the conductance with $R$, with the minimums 
(maximums) of $G$ located at the same values as the AF (FM)
correlations in Fig.~\ref{fig3}(a). Notice that for $U \geq 2$
the ratio between the maximum and the minimum values of $G$
becomes equal to 8 or larger. Again this very large ratio
should be detectable experimentally if a device with the
geometry of Fig.~\ref{fig1} and working at quarter-filling
could be fabricated.

\begin{figure}
\includegraphics[width=0.43\textwidth]{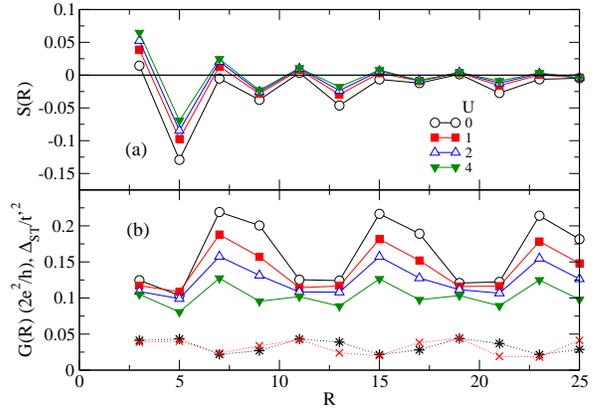}
\caption{(Color online) (a) Spin-spin correlations between the
impurities, and (b) conductance, as a function of the interimpurity
distance $R$, for $t^\prime=1$, at $n=0.5$, parameterized with $U$,
on the $L=120$ chain. In (b) the singlet-triplet energy 
$\Delta_{ST}/{t^\prime}^2$ is shown for $t^\prime=1$ (stars)
and $t^\prime=\sqrt{2}/2$ (crosses).}
\label{fig4}
\vspace{-0.5cm}
\end{figure}

In Fig.~\ref{fig4}(a), we show the spin-spin correlations between
impurities at $n=0.5$ for $t^\prime=1$. At this filling,
there is an additional important difference with the corresponding
results for $t^\prime=\sqrt{2}/2$ depicted in
Fig.~\ref{fig3}(a) which is that the $2k_F$ periodicity is now
clearly dominant. This behavior of $S(R)$ supports the idea that
these two values of $t^\prime$ belong to different
regimes.  On the other hand, the behavior of the conductance with
$R$ shown in Fig.~\ref{fig4}(b), in the noninteracting case
$U=0$, is similar to the one for $t^\prime=\sqrt{2}/2$, shown in
Fig.~\ref{fig3}(b), with a large $k_F$ component.
However, in the presence of interactions
on the chain the amplitude of $G(R)$ is {\em suppressed} by increasing
$U >0$ for $t^\prime=1$, while it seems {\em enhanced} by $U$ for
$t^\prime=\sqrt{2}/2$. This suppression of $G(R)$ would be expected 
in a Luttinger liquid since it is proportional to $K_\rho$ which
decreases with increasing $U$.\cite{kanefisher,qdhub} For
$t^\prime=1$, it can be also observed 
in Fig.~\ref{fig4}(b) that only for the largest value of the
interaction considered, $U=4$, a $2k_F$ periodicity becomes dominant.
Finally, the relation between the location of the maximum and 
minimum values of $G(R)$ with the ones of $S(R)$ observed in
Fig.~\ref{fig3}(b) is absent for small $U$ but it is recovered
for $U=4$.

These clearly different behaviors of $S(R)$ and $G(R)$ for the
two types of impurities considered have to be traced to the 
essential competition between the Kondo effect and RKKY 
interactions. For the noninteracting case, the Kondo temperature
may be estimated as\cite{hewson}
$T_K=t' \sqrt{U'/(2 t_0)} \exp{(-\pi t_0 U'/({8 t'}^2))}$
(in units of $k_B=1$) which gives $T_K=0.0026$ for
$t^\prime=\sqrt{2}/2$ and $T_K=0.086$ for $t^\prime=1$. That is,
if for the case of $t^\prime=\sqrt{2}/2$ it is reasonable to
assume that RKKY dominates over the Kondo effect, since $T_K$
for $t^\prime=1$ is $\approx 33$ times larger while the effective
RKKY interaction\cite{hirsch} ${\cal K} \sim {t^\prime}^2$ would be
only a factor of 2 larger, then it is plausible that the case of
$t^\prime=1$ belongs to the regime where the Kondo effect dominates
over the RKKY interaction. Particularly for the case of
$t^\prime=\sqrt{2}/2$ where RKKY can be assumed as dominant, it is
instructive to estimate ${\cal K}$ as the difference in ground state
energy between the singlet and triplet states. For this case
one gets that $\Delta_{ST}$ oscillates with $R$ with maximum
values of ${\cal K} \sim \Delta_{ST}\approx 0.04$ ($n=1$).
This value is larger than the above estimated $T_K$ for this
impurity. At $n=0.5$ the oscillation of $\Delta_{ST}$ with $R$
has period 8 as it can be observed in Fig.~\ref{fig4}(b).
Although these estimates are instructive and consistent with our
numerical results, one should bear in mind that the above
expression for $T_K$ is valid in the bulk limit while the
results for $\Delta_{ST}$ were obtained for the $L=120$ chain.
The strong suppression of $T_K$ due to the finite size of the
system found in Ref.~\onlinecite{simoncloud} may
not appear in our system due to the presence of the leads which
may provide room for the Kondo cloud to develop. We would also like
to notice that in all cases we examined the ground state is a 
singlet, which is consistent with the result that $S(R)$ is always
AF or at most weakly FM.

\begin{figure}
\includegraphics[width=0.43\textwidth]{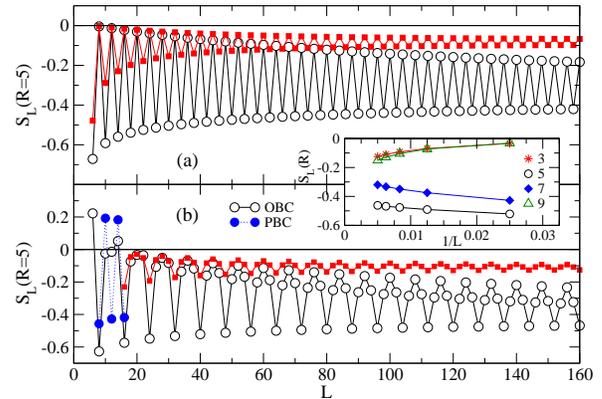}
\caption{(Color online) Spin-spin correlations between the impurities
at $R=5$, as a function of the total length $L$, for
$t^\prime=\sqrt{2}/2$ (circles), $t^\prime=1$ (squares), and
(a) $n=1$, (b) $n=0.5$. The inset shows the scaling of 
$S_L(R)$ with $L$, for $t^\prime=\sqrt{2}/2$, $n=0.5$, and 
for various values of $R$ as indicated.}
\label{fig5}
\vspace{-0.4cm}
\end{figure}

Finally in Fig.~\ref{fig5} we show the dependence of $S(R)$ at a
fixed interimpurity distance $R=5$ with the 
length $L$. The period of
$S_L(R=5)$ is 4 ($n=1$) and 8 ($n=0.5$) for the two types of
impurities considered, which correspond to wave vector
$k_F$. It is interesting to examine here the dependence of
these results with the boundary conditions (BC) of the
system.  Periodic BC (PBC) and antiperiodic BC (APBC) have,
separately, a period of 8 as with OBC. However, if for
each $L$ we adopt the BC with minimum energy, then a doubling
of the wave vector of the oscillation is obtained.  Using 
exact diagonalization we obtained that the energy is minimum for
PBC for $L=4m$ ($m$ integer) and for APBC
when $L=4m+2$. The minimum of energy for $L=4m+1$ and $L=4m+3$
falls alternatively on PBC or APBC depending on
$R$. The resulting $S_L(R=5)$ is shown in Fig.~\ref{fig5}(b).
This result is then another example of the effects caused by
OBC.\cite{kakashvili} The inset of Fig.~\ref{fig5}
shows a slow dependence of $S_L(R)$ with $L$ for various $R$.
Since, particularly for $t^\prime=\sqrt{2}/2$, $G(R)$ follows
the oscillation
of $S(R)$, it is quite likely then that an oscillation of the
conductance could be experimentally observed in a real device
for a fixed interdot distance as the length of the 1D leads is
varied.

\section{Conclusion}
\label{conclusion}

In conclusion, we have shown for the two-impurity Anderson model 
on finite chains, the presence of different behaviors of the
interimpurity magnetic correlations $S(R)$ for the two values of the 
coupling between the impurities and the chain, $t^\prime$. We 
suggested that these different behaviors indicate that the 
relative dominance between the Kondo effect and the
RKKY interaction can be tuned by the single parameter $t^\prime$. In
the case when RKKY dominates we found important oscillations in
the conductance with interimpurity distance,  $G(R)$, following
that of the spin-spin correlations. It would be 
tempting to relate these different behaviors of $S(R)$ and $G(R)$
to the presence of an unstable fixed point suggested by 
numerical renormalization group
calculations.\cite{tiam_main_res}
Although certainly such fixed point could not strictly appear in
a finite-size calculation some traces of its presence, for example
an admixture of the singlet and triplet states, could be detected.
However, we found that the ground state of our model 
is always a singlet consistently with dominating AF interimpurity
correlations. It would be interesting to explore other parameters
of our model to find such a critical point.\cite{affleck}

Finally, we have observed that the $k_F$ modulation of the $2 k_F$
oscillation of  $S(R)$ and $G(R)$ observed for noninteracting leads
is suppressed by an on-site electron repulsion on the chain. We hope
that some of the present results could be found in
a 1D realization of the device developed in \cite{craig},
which we believe could be built on a carbon nanotube with the QDs
defined with an appropriate array
of gates \cite{biercuk}.

\acknowledgments
One of us (SC) wishes to acknowledge partial financial support from
the Josefina Prats Foundation.


\begin{thebibliography}{}

\bibitem{tiam_history} C. Jayaprakash, H. R. Krishnamurty, and J. W.
     Wilkins, J. Appl. Phys.  {\bf 53}, 2142 (1982), and references
     therein.

\bibitem{hewson} A. J. Hewson, {\it The Kondo problem to heavy
      fermions}, (Cambridge University Press, Cambridge, UK, 1993).

\bibitem{tiam_main_res} B. A. Jones and C. M. Varma, Phys. Rev. Lett.
      {\bf 58}, 843 (1987); Phys. Rev. B {\bf 40}, 324 (1989).

\bibitem{hirsch} R. M. Fye, J. E. Hirsch, and D. J. Scalapino,
      Phys. Rev. B {\bf 35}, 4901 (1987).

\bibitem{goldhaber} D. Goldhaber-Gordon, H, Shtrikman, D. Mahalu,
    D. Abusch-Magder, U. Meirav, and M. A. Kastner, Nature
    {\bf 391}, 156 (1998); S. M. Cronenwett, T. H. Oosterkamp, and
      L. P. Kouwenhoven, Science {\bf 281}, 540 (1998); W. G. van der Wiel,
      S. De Franceschi, T. Fujisawa, J. M. Elzerman, S. Tarucha, and
      L. P. Kouwenhoven, Science {\bf 289}, 2105 (2000).

\bibitem{craig} N. J. Craig, J. M. Taylor, E. A. Lester, C. M. Marcus,
     M. P. Hanson, and A. C. Gossard, Science {\bf 304},
       565 (2004).

\bibitem{heersche} H. B. Heersche, Z. de Groot, J. A. Folk, L. P. 
     Kouwenhoven, H. S. J. van der Zant, A. A. Houck, J. Labaziewicz, and I. L. 
     Chuang, Phys. Rev. Lett.  {\bf 96}, 017205 (2006).

\bibitem{utsumi} Y. Utsumi, J. Martinek, P. Bruno, and H. Imamura,
      Phys. Rev. B {\bf 69}, 155320 (2004).

\bibitem{simoncloud} P. Simon, Phys. Rev. B {\bf 71}, 155319 (2005).

\bibitem{simon} P. Simon, R. L\'opez, and Y. Oreg, Phys. Rev. Lett.
        {\bf 94}, 086602 (2005).

\bibitem{bockrath} M. Bockrath, D. H. Cobden, P. L. McEuen, N. G. Chopra,
      A. Zettl, A. Thess, and  R. E. Smalley, Science {\bf 275},
      1922 (1997).

\bibitem{biercuk} M. J. Biercuk, S. Garaj, J. M. Chowm and C. M. Marcus,
      Nano Letters {\bf 5}, 1267 (2005).

\bibitem{egger-schoeller} R. Egger and H. Schoeller, Phys. Rev. B
       {\bf 54}, 16337 (1996).

\bibitem{dmrgrev} U. Schollw\"{o}ck, Rev.  Mod. Phys. {\bf 77}, 259
        (2005).

\bibitem{hallberg-egger} K. Hallberg and R. Egger, Phys. Rev. B
     {\bf 55}, R8646 (1997).

\bibitem{schollwock} U. Schollw\"ock, J. Phys. Soc. Jpn. {\bf 74}
        (Suppl.), 246 (2005), and references therein.

\bibitem{cazalilla} M. A. Cazalilla and J. B. Marston, Phys. Rev. Lett.
        {\bf 88}, 256403 (2002).

\bibitem{whitefeiguin} S. R. White and A. E. Feiguin, Phys. Rev. Lett.
        {\bf 93}, 076401 (2004).

\bibitem{alhassanieh} K. A. Al-Hassanieh, A. E. Feiguin, J. A. Riera,
       C. A. Busser, and E. Dagotto,
       Phys. Rev. B {\bf 73}, 195304 (2006).

\bibitem{schmitteckert} P. Schmitteckert, Phys. Rev. B {\bf 70}, 121302(R)
      (2004).

\bibitem{wingreen} N. S. Wingreen, A. P. Jauho, and Y.
        Meir, Phys. Rev. B {\bf 48}, 8487 (1993).

\bibitem{qdhub} S. Costamagna, C. J. Gazza, M. E. Torio, and J. A. Riera,
        Phys. Rev. B {\bf 74}, 195103 (2006).

\bibitem{lamnoimp} S. Costamagna and J. A. Riera,
        Phys. Rev. B {\bf 77}, 045302 (2008).

\bibitem{kanefisher} C. L. Kane and M. P. A.  Fisher, Phys. Rev. B
         {\bf 46}, 7268 (1992).

\bibitem{kakashvili} Effects of OBC on Luttinger liquid properties have 
      been studied by P. Kakashvili, H. Johannesson, and S. Eggert,
      Phys. Rev. B {\bf 74}, 085114 (2006).

\bibitem{affleck} See a further discussion of this problem in I. Affleck,
        A. W. W. Ludwig, and B. A. Jones, Phys. Rev. B {\bf 52}, 9528
        (1995).

\end{thebibliography}
\end{document}